# Is quantum entanglement invariant in special relativity?

D. Ahn*, H. J. Lee, S. W. Hwang†, and M. S. Kim‡

*Institute of Quantum Information Processing and Systems, University of Seoul, Seoul 130-743, Republic of Korea*

*† Department of Electronic and Computer Engineering, Korea University, Seoul 136-075, Republic of Korea*

*‡School of Mathematics and Physics, The Queen's University, Belfast BT7 1NN, United Kingdom*

**Abstract**

Quantum entanglements are of fundamental importance in quantum physics ranging from the quantum information processing to the physics of black hole. Here, we show that the quantum entanglement is not invariant in special relativity. This suggests that nearly all aspects of quantum information processing would be affected significantly when relativistic effects are considered because present schemes are based on the general assumption that entanglement is invariant. There should be additional protocols to compensate the variances of entanglement in quantum information processing. Furthermore, extending our results to general relativity may provide clues to the fate of the information contained in an entangled Hawking pair inside and outside the event horizon as black hole evaporates.

* e-mail: dahn@uoscc.uos.ac.kr

Quantum entanglement has novel features which are of considerable interest in quantum physics ranging from the quantum computation and quantum information processing [1-10] to the physics of black holes [11-15]. Quantum entanglement is an essential resource of quantum computation and quantum information to do information processing tasks impossible or quite difficult with classical methods [1]. Besides, it is suggested that the microscopic origin of the black hole entropy is the entanglement between Hawking particles inside and outside the event horizon [13, 14]. Current schemes of quantum computation and quantum information processing are mostly based on the conjecture that the entanglement is invariant. Especially, it has been generally assumed that the overall entanglement between all degrees of freedom, such as spin and momentum for the entire wave function, is invariant across different Lorentz frames [3-6]. At this stage, it would be interesting to study whether overall quantum entanglement is invariant in special relativity to discover the physical bounds of quantum information processing imposed by special relativity. Here, we show that the quantum entanglement is not invariant in special relativity, when each constituent particle has no definite momentum. This implies that the relativistic effects would affect nearly all schemes of quantum information processing significantly. For example, procedures based on the purification and stabilizer coding [1] will be frame dependent. Non-invariance of entanglement would also affect the joint measurements, and as a result, quantum cryptography using massive particles will be depending on the boost axis and reference frame. In general, there should be additional protocols to compensate the variances of entanglement, as one uses quantum error corrections for decoherence.

Extending the discussion of entanglement beyond non-relativistic theory brings new questions that need to be satisfactorily answered: 1) What is the covariant states corresponding to the two-particle entangled states? 2) Is the overall entanglement invariant and entanglement fidelity preserved in special relativity? The first question is now well understood [3-8,10], and it is known that the Lorentz transformation on the entangled state or the Bell state, $U(\Lambda)|\Psi\rangle_{AB}$, is the Wigner rotation in the new Lorentz frame [7,10]. The Lorentz transformation $\Lambda$ induces unitary transformation $U(\Lambda)$ on the state vectors in the Hilbert space,

$$\Psi \to U(\Lambda)\Psi , \tag{1}$$

where the transformation rule for the single particle state is given by [3-7,10]

$$U(\Lambda)\mid \vec{p},\sigma\rangle = \sqrt{\frac{(\Lambda p)^0}{p^0}}\sum_{\sigma'}D_{\sigma'\sigma}(\Omega_{\vec{p}})\mid \vec{p}_\Lambda,\sigma'\rangle . \tag{2}$$

Here $D(\Omega_{\vec{p}})$ is the Wigner representation for spin 1/2, $\Omega_{\vec{p}}$ is the Winger angle, $p^\mu=(\vec{p},p^0),(\Lambda p)^\mu=(\vec{p}_\Lambda,(\Lambda p)^0)$ with $\mu=1,2,3,0$, and $\sigma$ denotes spin of the particle. The Winger angle $\Omega_{\vec{p}}$ is determined by the boost speed and the original momentum $\vec{p}$ of the particle (2). The total state is given by

$$|\Psi\rangle_{AB} = \int d\vec{p}d\vec{q}f(\vec{p},\vec{q})|\vec{p},\vec{q}\rangle_{AB} \otimes |\Phi\rangle_{AB} \tag{3}$$

with $\int d\vec{p}d\vec{q}\mid f(\vec{p},\vec{q})\mid^2=1$. Here $|\vec{p},\vec{q}\rangle_{AB}$ is the momentum part and $|\Phi\rangle_{AB}$ is the spin part of the total state for Alice and Bob. Under the Lorentz transformation, the state seen by the observer in the moving frame becomes

$$U(\Lambda)|\Psi\rangle_{AB} = \int d\vec{p}d\vec{q}f(\vec{p},\vec{q})\sqrt{\frac{(\Lambda p)^0}{p^0}}\sqrt{\frac{(\Lambda q)^0}{q^0}}|\vec{p}_\Lambda,\vec{q}_\Lambda\rangle \otimes |\Phi(\vec{p},\vec{q})\rangle_{AB} , \tag{4}$$

where $|\Phi(\vec{p},\vec{q})\rangle_{AB} = \sum_{\sigma_A,\sigma_B,\sigma',\sigma''}D_{\sigma'\sigma_A}(\Omega_{\vec{p}})D_{\sigma''\sigma_B}(\Omega_{\vec{q}})|\sigma',\sigma''\rangle_{AB} . \tag{5}$

The Wigner representation for the boost in the x-direction and the momentum vector $\vec{p}=(p\cos\theta, p\sin\theta\cos\varphi, p\sin\theta\sin\varphi)$ is given by [7]

$$D(\Omega_{\vec{p}})=\begin{pmatrix}\cos(\Omega_{\vec{p}}/2)+i\sin(\Omega_{\vec{p}}/2)\cos(\varphi_{\vec{p}}) & -\sin(\Omega_{\vec{p}}/2)\sin(\varphi_{\vec{p}}) \\ \sin(\Omega_{\vec{p}}/2)\sin(\varphi_{\vec{p}}) & \cos(\Omega_{\vec{p}}/2)-i\sin(\Omega_{\vec{p}}/2)\cos(\varphi_{\vec{p}})\end{pmatrix}. \tag{6}$$

The second question whether the overall entanglement invariant and entanglement fidelity preserved in special relativity, is the main focus of this paper. If the overall entanglement is invariant, the transfer of entanglement between the spin and momentum degrees of freedom would be possible [6].

In order to find out whether quantum entanglement is invariant under Lorentz transformation, we need to quantify the degree of entanglement. There are several kinds of measures available for that, and among them are, the entanglement fidelity [16], the measure of entanglement [17, 18], and the concurrence [19]. The entanglement fidelity is the measure of how well the entanglement of the bipartite system is preserved, and can be applied to the entire wave function. The measure of entanglement and the concurrence are equivalent, but only give the degree of entanglement for the spin part. For the momentum state, it is difficult to define a suitable measure, and one needs to check the component of the momentum density matrix element, to see if it is separable [17]. In this paper, we are going to use the entanglement fidelity to check the total wave functions, the measure of entanglement for the spin part, and the components of the momentum density matrix to see if it separable.

In the following, we are going to show that: 1) entanglement fidelity is not preserved, 2) transfer of entanglement from the momentum to the spin is not possible, 3) for maximally entangled spin states, the measure of entanglement is degrading and spin states are no longer entangled when $\beta \to 1$, and 4) there is no transfer of entanglement from the spin into the momentum in the special case of $\beta \to 1$, where $\beta$ is the ratio of the boost and the light speed.

To check whether overall entanglement is preserved, we calculated the entanglement fidelity $F_{AB}$ for the moving frame, when the spins were maximally entangled and momentums were in the product state in the rest frame. Entanglement fidelity is a measure of how well the entanglement between $AB$ are maintained, and it is defined by [16]

$$F_{AB} = {}_{BA}\langle\Psi\big|\big(U(\Lambda)\big|\Psi\rangle_{AB\ BA}\langle\Psi\big|U^{-1}(\Lambda)\ \big)\big|\Psi\rangle_{AB} = |{}_{BA}\langle\Psi|U(\Lambda)|\Psi\rangle_{AB}|^2, \quad (8)$$

where

$${}_{BA}\langle\Psi|U(\Lambda)|\Psi\rangle_{AB} = \int d\vec{p}d\vec{q}\sqrt{\frac{(\Lambda p)^0}{p^0}}\sqrt{\frac{(\Lambda q)^0}{q^0}}f*(\Lambda\vec{p},\Lambda\vec{q})f(\vec{p},\vec{q})\otimes {}_{BA}\langle\Psi\big\|\Psi(\vec{p},\vec{q})\rangle_{AB}$$

$$= \int d\vec{p}d\vec{q}\sqrt{\frac{(\Lambda p)^0}{p^0}}\sqrt{\frac{(\Lambda q)^0}{q^0}}f*(\Lambda\vec{p},\Lambda\vec{q})f(\vec{p},\vec{q})\cos(\Omega_{\vec{p}}/2)\cos(\Omega_{\vec{q}}/2). \quad (9)$$

Here we assumed that the distribution function is of the form $f(\vec{p},\vec{q}) = \sqrt{N\exp(-\vec{p}^2/\Delta)}\sqrt{N\exp(-\vec{p}^2/\Delta)}$, when the initial spin state is maximally entangled. For $0 < \beta \le 1$,

$$\frac{(\Lambda p)^0}{p^0}N\exp(-(\Lambda\vec{p})^2/\Delta) = N\exp(-\vec{p}^2/\Delta)\frac{(\Lambda p)^0}{p^0}\exp(-\frac{(p^0)^2}{\Delta}((\Lambda p^0/p^0)^2 - 1)$$

$$< N\exp(-\vec{p}^2/\Delta), \text{ since } \Lambda p^0/p^0 > 1 \text{ for } \beta > 0. \quad (10)$$

Thus, we have $F_{AM} = |{}_{BA}\langle\Psi|U(\Lambda)|\Psi\rangle_{AB}|^2$

$$< \int d\vec{p}d\vec{q}f*(\vec{p},\vec{q})f(\vec{p},\vec{q})\cos(\Omega_{\vec{p}}/2)\cos(\Omega_{\vec{q}}/2)$$

$$< 1. \quad (11)$$

We assumed that the x-axis is the boost direction, and the entangled states in the rest frame are formed by the eigenstates of the spin in the z-direction. It is shown that the entanglement fidelity is less than 1 in the moving frame. This implies that entanglement fidelity is not preserved in special relativity.

Next, we consider if it is possible to transfer the entanglement from the momentum to the spin under the Lorentz transformation. We assume that the state in the rest frame was maximally entangled for the momentum, but in the product state for the spins. By writing $U(\Lambda)|\Psi\rangle_{AB}$ as a density matrix and tracing over the momentum degrees of freedom in the moving frame, we obtain the reduced density matrix for the spin as

$$\rho_{AB} = \int d\vec{p}\,''d\vec{q}\,''\langle \vec{p}\,'',\vec{q}\,''|U(\Lambda)|\Psi\rangle_{AB}{}_{BA}\langle\Psi|U^{-1}(\Lambda)|\vec{q}\,'',\vec{p}\,''\rangle$$

$$= \int d\vec{p}d\vec{q} \mid f(\vec{p},\vec{q}) \mid^2 \mid \Phi(\vec{p},\vec{q})\rangle_{AB}{}_{BA}\langle\Phi(\vec{p},\vec{q})\mid. \qquad (12)$$

Here, we have used the relativistically invariant delta function $p^0\delta(\vec{p}-\vec{p}_{\Lambda^{-1}}'') = (\Lambda p)^0\delta(\vec{p}_\Lambda - \vec{p}'')$ and Lorentz invariant volume element $d\vec{p}/p^0 = d\vec{p}_{\Lambda^{-1}}''/(p'')^0$. Non-vanishing part of the reduced density matrix for the momentum, seen by the moving observer becomes

$$\rho_{AB} = \int d\vec{p}d\vec{q} \mid f(\vec{p},\vec{q})\mid^2 \begin{pmatrix} \mid a\mid^2 & 0 & 0 & ad* \\ 0 & \mid b\mid^2 & bc* & 0 \\ 0 & b*c & \mid c\mid^2 & 0 \\ a*d & 0 & 0 & \mid d\mid^2 \end{pmatrix}, \qquad (13)$$

for initially unentangled spin state in the rest frame $|\Phi\rangle_{AB} = |\uparrow,\uparrow\rangle_{AB}$, and $\mid a\mid^2 + \mid b\mid^2 + \mid c\mid^2 + \mid d\mid^2 = 1$. Details of $a,b,c,d$ are given in the Appendix. Moreover, we have

$$\left(<\mid a\mid^2>\right)\left(<\mid d\mid^2>\right) = \left(<\mid b\mid^2>\right)\left(<\mid c\mid^2>\right) \quad \text{and} \quad \mid<bc*>\mid \ \geq \ \mid<ad*>\mid \qquad (14)$$

with $<\bullet> = \int d\vec{p}d\vec{q}\mid f(\vec{p},\vec{q})\mid^2 \bullet$ for initially maximum entangled momentum state. The density matrix $\rho_{AB}$ is inseparable (or A and B are entangled), if and only if the partial transposition of $\rho_{AB}$, $\sigma_{AB} = \rho_{AB}^{T_2}$ has any negative eigenvalues [17,18]. The condition that at least one of the eigenvalues for the partial transposition $\sigma_{AB}$ is negative, is given by

$$|<ad*>|^2 \quad > \quad \left(<|b|^2>\right)\left(<|c|^2>\right), \tag{15}$$

or

$$|<bc*>|^2 \quad > \quad \left(<|a|^2>\right)\left(<|d|^2>\right) \tag{16}$$

For maximum entangled momentum state, the typical distribution function is of the form $|f(\vec{p},\vec{q})|^2 = |g(\vec{p})|^2 \, \delta(\vec{p} \pm \vec{q})$, and one can show that non-vanishing elements of the reduced density matrix are as given by equation (13) and to derive equations (14) to (16) after some mathematical manipulations. It is not too difficult to show that none of above inequalities is true mathematically by using equation (14) and Cauchy-Schwartz inequality. So there is no transfer of entanglement from the momentum degrees of freedom to the spins under the Lorentz transformation.

In order to find out if momentum state remains maximally entangled, we have taken the partial trace over the spin degrees of freedom for the density matrix. It is found that each element of the momentum density matrix is weighted by the factor of the form $F(\vec{p},\vec{p}',\vec{q},\vec{q}',\beta) + iG(\vec{p},\vec{p}',\vec{q},\vec{q}',\beta)$. It is very likely that the momentum state is no longer maximally entangled under the boost, because the weighting factor causes the oscillations among the matrix elements. Since we started from maximally entangled momentum state, if the overall entanglement is invariant, there should have been a transfer of entanglement to the spin degrees of freedom.

We have done the similar analysis for the case of maximally entangled spin in the rest frame. It is assumed that the initial momentum state was in the product state, i.e., $f(\vec{p},\vec{q}) = f_1(\vec{p})f_2(\vec{q})$. For the initial spin state $|\Phi\rangle_{AB} = \frac{1}{\sqrt{2}}\left(|\uparrow\uparrow\rangle_{AB} + |\downarrow\downarrow\rangle_{AB}\right)$, we obtain the following eigenvalues for the partial transposition $\sigma_{AB}$:

$$\lambda = (1-2A)/2, (1-2B)/2, (1-2C)/2, (1-2D)/2, \tag{17}$$

where $A, B, C, D$ are non-vanishing elements of the density operator (Appendix):

$$\rho_{AB} = \begin{pmatrix} (A+D)/2 & 0 & 0 & (A-D)/2 \\ 0 & (B+C)/2 & -(B-C)/2 & 0 \\ 0 & -(B-C)/2 & (B+C)/2 & 0 \\ (A-D)/2 & 0 & 0 & (A+D))/2 \end{pmatrix}, A+B+C+D=1.$$

In the non-relativistic limit $\beta = v/c \to 0$, we obtain $A=1, B=C=D=0$, and one of the eigenvalues of partial transposition matrix is negative, which is equal to $-1/2$. The measure of entanglement [18] becomes $E(\rho_{AB}) = 1$, implying the maximal entanglement for the spin. On the other hand, in the ultra-relativistic limit $\beta \to 1$, we get

$$A = 1 - \eta + 3\eta^2/8, B = D = \eta/2 - \eta^2/4, C = \eta^2/8 \qquad (18)$$

where $\eta = \int_0^\pi \sin(\theta)\sin^2(\Omega_p/2)$ and becomes 1 because $\Omega_{\bar{p}} \to \theta$ for $p^0/m >> 1$. Then the none of the eigenvalues for the partial transposition matrix is negaitve and the spins are no longer entangled. In general, the measure of entanglement is monotonically decreasing function of $\beta$.

In order to check if there is a transfer of entanglement from the spin to the momentum, we have again taken the partial trace over the spin degrees of freedom. After some mathematical manipulations, non-vanishing elements of the reduced density matrix become

$$\rho_{AB}' = \sum_{\sigma,\sigma'} \langle \sigma, \sigma' | U(\Lambda) | \Psi \rangle_{AB} \, _{BA}\langle \Psi | U^{-1}(\Lambda) | \sigma, \sigma' \rangle,$$

and $\quad \langle \vec{p}, \vec{q} | \rho_{AB}' | \vec{p}', \vec{q}' \rangle \to \langle \vec{p} | \rho_A | \vec{p}' \rangle \otimes \langle \vec{q} | \rho_B | \vec{q}' \rangle, \qquad (19)$

for $\beta \rightarrow 1$ and $p^0/m >> 1$, which shows that momentum state remains in the product state, and indicates there is no transfer of entanglement to the momentum degrees from the spin under Lorentz transformation.

Therefore, we conclude that the overall entanglement is not invariant. Since we have proved that the entanglement is not invariant, the above observation of the degrading of spin entanglement does not necessarily mean that the entanglement in spin can be transferred to the momentum, or vice versa. It was shown [6] that the entanglement just between the spins of a pair of particles is not invariant, and it was concluded [6] that under a Lorentz transformation, the initial entanglement of just the spin degrees of freedom can be transferred into an entanglement between both the spin and the momentum degrees of freedom, or vice versa. That conclusion was based on the conjecture that local unitary operations leave the entanglement invariant [3]. However, when the total state has no definite momentum, the results of the Lorentz transformation may not be necessarily represented by local unitary operation on the original state, in general, because the Lorentz transformation involves both spin and momentum, and one needs to integrate the transformed states over the momentum, weighted by the distribution function, which destroy the unitary properties. Historically, this is somewhat similar to the case of the finding of parity non-conservation in connection with the $\vartheta - \tau$ puzzle and the beta-decay in 1956-1957 by Lee, Yang, and Wu, when most people took for granted that the parity is conserved [20]. As a potential implications of the effects of Lorentz transformation on the information transfer between the rest frame and the moving frame, we study the quantum correlations seen by the moving observer. Since any Hermitian operator defines an observables, one could define the normalized relativistic observable $\hat{a}$ as [2,21],

$$\hat{a} = \frac{[\sqrt{1-\beta^2}(\vec{a}-\vec{e}(\vec{a}\bullet\vec{e}))+\vec{e}(\vec{a}\bullet\vec{e})]}{\sqrt{1+\beta^2((\vec{e}\bullet\vec{a})^2-1)}}\bullet\vec{\sigma}, \qquad (20)$$

where we normalized the relativistic observable by the absolute value of its eigenvalue. Here $\vec{a}$ is the direction in the moving frame and $\vec{\sigma}$ is the Pauli matrix. Normalized relativistic observable $\hat{a}$ is related to the Pauli-Lubanski pseudo -vector, which is known to be a relativistic invariant operator corresponding to spin. It should be noted that ordinary Pauli matrix $\vec{\sigma}$ and $\vec{a}\bullet\vec{\sigma}$ are not, in general, relativistic observables [22,23].

It is straightforward to calculate the classical correlation $\langle\hat{a}\hat{b}\rangle_{classical}$, when the moving observer is receding (approaching) from (to) the rest frame with the speed of light, and is given by

$$\langle\hat{a}\hat{b}\rangle_{classical} = \frac{\vec{a}\bullet\vec{e}}{|\vec{a}\bullet\vec{e}|}\bullet\frac{\vec{b}\bullet\vec{e}}{|\vec{b}\bullet\vec{e}|} = \pm 1. \qquad (21)$$

In the calculation of corresponding quantum correlations, we assumed that both spin and momentum states are maximally entangled. Then the quantum correlation $\langle\hat{a}\hat{b}\rangle_{quantum}$ calculated in the moving frame becomes,

$$\langle\hat{a}\hat{b}\rangle_{quantum} \rightarrow \frac{a_x}{|a_x|}\frac{b_x}{|b_x|}\int d\vec{p}d\vec{q}\,|f(\vec{p},\vec{q})|^2\left(X^2-Y^2-Z^2+W^2\right)\leq\pm 1 \;\; \text{as}\,\beta\rightarrow 1, \quad (22)$$

for general momentum $\vec{p}$ and the boost in the x-axis. Here $X,Y,Z,W$ are given in the **Appedix**. We also have the following relation, $2\sin^2\theta\sin^2\varphi-1\leq X^2-Y^2-Z^2+W^2\leq 1$. By comparing eqs. (21) and (22), one can see that the quantum correlation approaches to the classical correlation, when the speed of the moving observer approaches the speed of light, and in both cases, the information in the vertical direction to the boost axis is lost.

Finally, we would like to mention that the extension of our result to general relativity could also have practical implications on the physics of black hole. Even though much of black hole physics is still in *terra incognita*, it is generally accepted that the black hole has entropy [10,11]. Its microscopic origin is suggested to be the entanglement between particles entangled inside and outside of the event horizon [13,14]. Whether the information contained in the black hole re-emerge as the black hole evaporates is an important issue in cosmology [15]. Extending relativistic entanglement to non-inertial frame may provide a way to find clues to the fate of the information contained in an entangled Hawking pair.

## Acknowledgements

This work was supported by the Korean Ministry of Science and Technology through the Creative Research Initiatives Program under Contract No. M1-0016-00-0008. D. A. also thanks H. Ahn and T. Yim for encouragements.

## Appendix

### (1) Detailed expression for $a, b, c, d$ :

$$a(\vec{p}, \vec{q}) = \cos(\Omega_{\vec{p}}/2)\cos(\Omega_{\vec{q}}/2) - \sin(\Omega_{\vec{p}}/2)\sin(\Omega_{\vec{q}}/2)\cos(\varphi_{\vec{p}})\cos(\varphi_{\vec{q}})$$
$$+ i\left(\sin(\Omega_{\vec{p}}/2)\cos(\Omega_{\vec{q}}/2)\cos(\varphi_{\vec{p}}) + \cos(\Omega_{\vec{p}}/2)\sin(\Omega_{\vec{q}}/2)\cos(\varphi_{\vec{q}})\right), \quad (A1)$$

$$b(\vec{p}, \vec{q}) = \cos(\Omega_{\vec{p}}/2)\sin(\Omega_{\vec{q}}/2)\sin(\varphi_{\vec{q}}) + i\sin(\Omega_{\vec{p}}/2)\sin(\Omega_{\vec{q}}/2)\cos(\varphi_{\vec{p}})\sin(\varphi_{\vec{q}}), \quad (A2)$$

$$c(\vec{p}, \vec{q}) = \sin(\Omega_{\vec{p}}/2)\cos(\Omega_{\vec{q}}/2)\sin(\varphi_{\vec{p}}) + i\sin(\Omega_{\vec{p}}/2)\sin(\Omega_{\vec{q}}/2)\sin(\varphi_{\vec{p}})\cos(\varphi_{\vec{q}}), \quad (A3)$$

and $d(\vec{p}, \vec{q}) = \sin(\Omega_{\vec{p}}/2)\sin(\Omega_{\vec{q}}/2)\sin(\varphi_{\vec{p}})\sin(\varphi_{\vec{q}})$. $\quad (A4)$

### (2) Mathematical expressions for $A, B, C, D$ :

For maximally entangled spin state $\left|\Phi\right\rangle_{AB} = \frac{1}{\sqrt{2}}\left(\left|\uparrow\uparrow\right\rangle_{AB} + \left|\downarrow\downarrow\right\rangle_{AB}\right)$, we obtain

$$A = \int d\vec{p}d\vec{q} \mid f(\vec{p},\vec{q})\mid^2 \begin{pmatrix} \cos^2(\Omega_{\vec{p}}/2)\cos^2(\Omega_{\vec{q}}/2) \\ + \sin^2(\Omega_{\vec{p}}/2)\sin^2(\Omega_{\vec{q}}/2)\cos^2(\varphi_{\vec{p}}+\varphi_{\vec{q}}) \end{pmatrix}, \qquad (A5)$$

$$B = \int d\vec{p}d\vec{q} \mid f(\vec{p},\vec{q})\mid^2 \begin{pmatrix} \sin^2(\Omega_{\vec{p}}/2)\cos^2(\Omega_{\vec{q}}/2)\sin^2(\varphi_{\vec{p}}) \\ + \cos^2(\Omega_{\vec{p}}/2)\sin^2(\Omega_{\vec{q}}/2)\sin^2(\varphi_{\vec{q}}) \end{pmatrix}, \qquad (A6)$$

$$C = \int d\vec{p}d\vec{q} \mid f(\vec{p},\vec{q})\mid^2 \left(\sin^2(\Omega_{\vec{p}}/2)\sin^2(\Omega_{\vec{q}}/2)\sin^2(\varphi_{\vec{p}}+\varphi_{\vec{q}})\right), \qquad (A7)$$

and $$D = \int d\vec{p}d\vec{q} \mid f(\vec{p},\vec{q})\mid^2 \begin{pmatrix} \sin^2(\Omega_{\vec{p}}/2)\cos^2(\Omega_{\vec{q}}/2)\cos^2(\varphi_{\vec{p}}) \\ + \cos^2(\Omega_{\vec{p}}/2)\sin^2(\Omega_{\vec{q}}/2)\cos^2(\varphi_{\vec{q}}) \end{pmatrix}, \qquad (A8)$$

from equations (3)-(6) and equation (12).

## (3) Expressions for $X, Y, Z, W$:

$$X = \cos((\Omega_{\vec{p}}+\Omega_{-\vec{p}})/2)\sin^2(\varphi_{\vec{p}}) + \cos((\Omega_{\vec{p}}-\Omega_{-\vec{p}})/2)\cos^2(\varphi_{\vec{p}}), \qquad (A9)$$

$$Y = \sin((\Omega_{\vec{p}}+\Omega_{-\vec{p}})/2)\sin(\varphi_{\vec{p}}), \qquad (A10)$$

$$Z = \sin((\Omega_{\vec{p}}-\Omega_{-\vec{p}})/2)\cos(\varphi_{\vec{p}}), \qquad (A11)$$

and $W = -\cos((\Omega_{\vec{p}}+\Omega_{-\vec{p}})/2)\sin(\varphi_{\vec{p}})\cos(\varphi_{\vec{p}}) + \cos((\Omega_{\vec{p}}-\Omega_{-\vec{p}})/2)\sin(\varphi_{\vec{p}})\cos(\varphi_{\vec{p}})$.